\documentclass[11pt]{article}
\usepackage{graphicx}  

\begin{document}
\begin{titlepage}
\title{
\begin{flushright}\begin{small}
LAPTH-1286/08
\end{small}\end{flushright}
\vspace{2cm} Sigma-model approaches to exact solutions in higher-dimensional
gravity and supergravity\thanks{Talk presented at the WE Heraeus Seminar on 
Models of Gravity in Higher Dimensions: From Theory to Experimental Search, 
Bremen, 25-29.8.2008}} \author{G\'erard Cl\'ement
\thanks{Email: gclement@lapp.in2p3.fr} \\ \\ \small Laboratoire de
Physique Th\'eorique LAPTH (CNRS), \\ \small B.P.110, F-74941
Annecy-le-Vieux cedex, France} \date{4 November 2008} \maketitle

\abstract{Classical gravitating field theories reduced to three dimensions
admit manifest gauge invariances and hidden symmetries, which
together make up the invariance group $G$ of the theory. If this group
is large enough, the target space is a symmetric space $G/H$. New
solutions may be generated by the action of invariance transformations
on a seed solution. Another application is the construction of
multicenter solutions from null geodesics of the target space.
After a general introduction on this sigma-model approach, I will
discuss the case of five-dimensional gravity, with invariance group
$SL(3,R)$, and minimal five-dimensional supergravity, with invariance
group $G_{2(2)}$. I will also describe recent attempts at the generation
of new charged rotating black rings.}

\end{titlepage}
\setcounter{page}{2}
\section{Introduction}
\label{intro}
An important field in classical general relativity in four or higher 
dimensions is the search for exact solutions describing black objects 
\cite{ER08} (black holes, black rings, etc.). While various techniques to 
derive particular solutions are available, it may be very difficult to 
find the general solution (of a given type), i.e. with
the maximum number of independent physical parameters.

For example, the most general black ring \cite{ER06} solution to minimal
five-dimensional supergravity (five-dimensional Einstein-Maxwell theory with
a Chern-Simons term, hereafter abbreviated as EM5) should have five independent
parameters: a mass
$M$, two angular momenta $J_{\psi}$ and $J_{\phi}$, an electric charge $Q$,
and a magnetic dipole charge $D$. At present, only black ring solutions with
three independent parameters are known:
\begin{itemize}
\item The non-supersymmetric black rings constructed in \cite{EEF05}
(Sect. 4). These apparently have five parameters, but two constraints to
ensure the absence of Dirac-Misner strings and of conical defects leave only
three independent parameters.

\smallskip
\item The black rings with two angular momenta constructed in \cite{pomsen}
(three parameters $M$, $J_{\psi}$ and $J_{\phi}$).

\smallskip
\item The charged black rings with two angular momenta constructed in
\cite{g2} apparently have four independent parameters $M$, $J_{\psi}$,
$J_{\phi}$, and $Q$, but are singular due to the presence of Dirac-Misner
strings.
\end{itemize}

A solution to the problem of generating exact solutions is provided by the
sigma-model approach. Usually the solutions of interest are stationary
(Killing vector $\partial_t$) and rotationally symmetric (Killing vector
$\partial_{\varphi}$), allowing e.g. the dimensional reduction
of a five-dimensional gravitating field theory to a three-dimensional 
gravitating sigma model. If one is lucky enough,
the target space $\cal T$ of this sigma model is a symmetric space, or coset
$G/H$, where $G$ is the group of isometries of $\cal T$ ($G_{2(2)}$ for EM5), 
and $H \subset G$ the local isotropy subgroup ($SL(2,R) \times SL(2,R)$ for 
EM5). It is then possible to generate new solutions by applying group
transformations to the coset representative of a seed solution. Another
application is the construction of multicenter solutions as totally geodesic
submanifolds of the target space.

In the next section I will give a brief overview of the sigma-model approach.
This will be followed by the application to five-dimensional general
relativity (E5) in section \ref{sec3}, and to minimal five-dimensional
supergravity (EM5), including recent progress in the generation of charged
rotating black rings, in section \ref{sec4}.

\section{Pedestrian overview of the sigma-model approach}
\label{sec2}
\setcounter{equation}{0}

\subsection{The SL(2,R)/SO(2) sigma model}\label{sub21}
There are many good reviews on the subject of gravitating sigma
models \cite{kra,BM}. Let me introduce the basic ideas on the
well-known example of four-dimensional vacuum gravity (E4) with one
Killing vector $\partial_t$ \cite{ernst}. The standard Kaluza-Klein
(KK) reduction from four to three dimensions leads to the line
element
\begin{equation}\label{st4}
ds^2 =  -f(dt + a_idx^i)^2 + f^{-1}h_{ij}dx^idx^j\,,
\end{equation}
where the gravitational potential $f$ and the three-dimensional reduced metric
$h_{ij}$ ($i=1,2,3$) depend only on the $x^i$. The reduced vacuum Einstein
equations ${R_{\mu}}^{\nu}=0$ split in three components:
\begin{itemize}
\item
a vector component ${R_0}^i = 0$, written symbolically as
\begin{equation}
\nabla\wedge(f^2\nabla\wedge\vec{a})=0\,;
\end{equation}
this is solved by the duality equation
\begin{equation}\label{duale4}
\nabla\wedge\vec{a} = f^{-2}\nabla\omega\,,
\end{equation}
which enables us to trade the Kaluza-Klein vector potential $\vec{a}$ for a
scalar potential $\omega$, solving identically the field equation
\begin{equation}\label{om}
\nabla(f^{-2}\nabla\omega)=0\,;
\end{equation}
\item
a scalar component $R_{00} = 0$, leading to the field equation
\begin{equation}\label{f}
f\nabla^2f = (\nabla f)^2 - (\nabla\omega)^2\,;
\end{equation}
\item
and a tensor component $R^{ij} = 0$, leading to the field
equation
\begin{equation}\label{eine4}
R_{(3)ij}(h) = \frac1{2f^2} \bigg(\partial_i f\partial_j f +  \partial_i\omega
\partial_j\omega\bigg) \,.
\end{equation}
\end{itemize}
The equations (\ref{om}), (\ref{f}) and (\ref{eine4}) derive from the
reduced action
\begin{equation}
S_{(3)} = \int d^3x\sqrt{|h|}\bigg[-R_{(3)}(h) + G_{AB}(X)\partial_iX^A
\partial_jX^Bh^{ij}\bigg]\,,
\end{equation}
which defines the three-dimensional (coordinates $x^i$, metric $h_{ij}$)
gravitating sigma model for the two-dimensional target space (also called
potential space \cite{kra}) $\cal T$
(coordinates $X^A$, metric $G_{AB}$) with the line element
\begin{equation}\label{tare4}
dS^2 \equiv G_{AB}dX^AdX^B = \frac1{2f^2}(df^2+d\omega^2)\,.
\end{equation}

Two obvious isometries of the target space line element (\ref{tare4})
follow from the definitions of the the potentials $f$ and $\omega$. The
line element (\ref{st4}) is form-invariant under time rescalings (a relic
of the original spacetime invariance under diffeomorphisms, or gauge
invariance) $t \to \alpha^{-1}t$, provided the potentials scale as
$f \to \alpha^2 f$, $\omega \to \alpha^2\omega$. These rescalings are
generated by the Killing vector of (\ref{tare4})
\begin{equation}
M = 2(f\partial_f + \omega\partial_{\omega})\,.
\end{equation}
Also, the duality equation (\ref{duale4}) defines the potential $\omega$
only up to translations, generated by the Killing vector
\begin{equation}
N = \partial_{\omega}\,.
\end{equation}
Besides these {\em manifest} symmetries, the line element (\ref{tare4})
also admits the {\em hidden} symmetry under infinitesimal transformations
\begin{equation}
L = (\omega^2+f^2)\partial_{\omega} + 2\omega f\partial_f\,.
\end{equation}
These three Killing vectors generate the Lie algebra
\begin{eqnarray}
\left[M,N\right] &=& -2N \,,\nonumber\\
\left[M,L\right] &=& 2L \,,\\
\left[N,L\right] &=& M \,,\nonumber
\end{eqnarray}
which we recognize as Lie[$SL(2,R)$].

The target space $\cal T$ is the symmetric space $SL(2,R)/SO(2)$. 
A familiar representation of this coset is in terms of
the complex Ernst potential
\begin{equation}
{\cal E} = f + i\omega\,,
\end{equation}
leading to the well-known Ernst equations \cite{ernst}. Less well
known, but better suited for generalization to other sigma models,
is the representation in terms of a symmetrical $2\times2$ matrix
\begin{equation}
{\cal M} = \left(\begin{array}{cc}
f + f^{-1}\omega^2 & -f^{-1}\omega \\
-f^{-1}\omega &  f^{-1}
\end{array}\right)\,,
\end{equation}
which allows to write the target space metric (\ref{tare4}) as
\begin{equation}\label{tarmet}
dS^2 = \frac14{\rm Tr}({\cal M}^{-1}d{\cal M}
{\cal M}^{-1}d{\cal M})\,,
\end{equation}
and the vacuum Einstein equations (\ref{om}), (\ref{f}) and (\ref{eine4}) as
\begin{eqnarray}
\nabla({\cal M}^{-1}\nabla{\cal M}) &=& 0\,, \label{divJ}\\
R_{(3)ij}(h) &=& \frac14{\rm Tr}({\cal M}^{-1}\partial_i{\cal M}
{\cal M}^{-1}\partial_j{\cal M})\,. \label{Rij}
\end{eqnarray}

Other examples of gravitating sigma models obtained by dimensional reduction
to three dimensions include:
\begin{itemize}
\item $D=4$ Einstein-Maxwell theory (EM4) with one Killing vector, leading
to the coset \cite{eh,ernst,neukra,kin73,mazur}
$${\cal T} = SU(2,1)/S[U(2)\times U(1)].$$
\item $D=p+3$ vacuum Einstein gravity (E$D$) with $p$ commuting Killing
vectors, leading to the coset \cite{maison}
$${\cal T} = SL(p+1,R)/SO(2,p-1).$$
\item $D=4$ Einstein-Maxwell-dilaton-axion theory (EMDA) with one Killing
vector, for which the coset is \cite{galtsov94}
$${\cal T} = Sp(4,R)/U(2).$$
Note that this example is intimately related to the preceding, as EMDA
has been shown to be a sector of $D=6$ vacuum Einstein gravity \cite{bremda}
(see also \cite{cedric}).
\item $D=11$ supergravity with 8 commuting Killing vectors, leading to the
coset
\cite{julia81}
$${\cal T} = E_{8(8)}/SO(16).$$
\end{itemize}

\subsection{Applications of gravitating sigma models}
\subsubsection{Generation of new solutions.}
It is clear that a global group transformation
\begin{equation}\label{transf}
{\cal M}(x) \to {\cal M}'(x) = P^T{\cal M}(x)P \qquad (P \in G)
\end{equation}
leaves invariant the field equations (\ref{divJ}) and the right-hand side of
(\ref{Rij}), and so preserves the reduced three-dimensional metric $h_{ij}$.
After reconstructing from the matrix ${\cal M}'(x)$ the local fields of the
full (non-reduced) theory, this transformation thus leads to a new solution
of this theory. Asymptotic flatness will be preserved {\em iff}
$P \in H_{\infty}$ (the isotropy subgroup at infinity).

Consider again the example of E4 with the Minkowkian signature $-+++$. For
an asymptotically flat solution, $f(\infty) = 1$, and $\omega=0$, leading to
\begin{equation}
{\cal M}_{\infty} = \left(\begin{array}{cc}
1 & 0 \\ 0 & 1
\end{array}\right)\,.
\end{equation}
In this case $H_{\infty}$ is the rotation group $SO(2)$, with only one element
which may be parametrized as
\begin{equation}
P = \left(\begin{array}{cc}
\cos\alpha & -\sin\alpha \\ \sin\alpha & \cos\alpha
\end{array}\right)\,.
\end{equation}
Applying the corresponding transformation (\ref{transf}) to the Schwarzschild
solution $f = 1 - 2m/r$, $\omega = 0$, we find that asymptotically
$\omega' \sim2m\sin2\alpha\cdot r^{-1}$, which dualizes to $a_{\varphi}' \sim
-2m\sin2\alpha\cos\theta$, corresponding to the Schwarzschild-NUT solution.

In the case of Einstein-Maxwell theory (EM4), similar three-parameter
group transformations applied to the Kerr solution reproduce the well-known
dyonic Kerr-Newman NUT solutions. More interestingly, we shall see in the
next section that, in the case of E5, the same approach was used by Rasheed
\cite{rasheed} to generate previously unknown rotating black string solutions.
\subsubsection{Multicenter solutions.}
In the case of precise equality between mass $m$ and electric charge $q$
(in gravitational units), Newtonian attraction and
Coulombian repulsion exactly compensate so that stationary multiparticle
systems are possible. This non-relativistic argument carries over to
relativistic Einstein-Maxwell theory, in the framework of which static
multicenter solutions were first constructed by Papapetrou and Majumdar
\cite{papa}. These were later generalized to stationary multicenter solutions
by Neugebauer, Perj\`es, and Israel and Wilson \cite{neuge,IWP}. I first
showed in 1986 how to construct such multicenter solutions from null
geodesics of the target space for five-dimensional gravity (E5) \cite{spat}.
This method, which was later generalized to the case of other gravitating
sigma models \cite{bps}, shall be described in section \ref{sec3}.
\subsubsection{Geroch group.}
Often the number of commuting Killing vectors of the spacetime is larger than
$D-3$, i.e. not all the Killing vectors are used for reduction to three
dimensions. For instance, a particularly interesting subclass of stationary
solutions of four-dimensional Einstein-Maxwell theory is that of stationary
axisymmetric solutions, with two Killing vectors $\partial_t$ and
$\partial_{\varphi}$.
In that case, one can combine transformations of the global symmetry group
$G$ with transformations in the hyperplane of Killing vectors, leading to the
so-called infinite-dimensional ``Geroch group'' \cite{geroch}. These
transformations allow in principle the generation of all solutions of the
stationary axisymmetric Einstein--Maxwell problem, which is thus completely
integrable. Practically, this generation of stationary axisymmetric solutions
is achieved via inverse--scattering transform methods \cite{maison78}.
It is only recently that finite Geroch transformations allowing the direct
generation of rotating solutions from static solutions were found
\cite{kerr,giusax}. I shall describe these approaches in subsections
\ref{subgiusax} and \ref{subkerr}.

\section{The case of five-dimensional general relativity (E5)}
\label{sec3}
\setcounter{equation}{0}
\subsection{The Maison approach}
The reduction of five-dimensional vacuum gravity to three dimensions and
the determination of the isometries of the resulting target space were first
achieved by Neugebauer \cite{neuge}, who found an eight-parameter symmetry
group. Neugebauer's approach was subsequently applied by Matos \cite{matos} to
solution generation. However it is less transparent than Maison's approach
\cite{maison}, which has the advantage of being manifestly covariant under
$GL(2,R)$ transformations in the plane of the two Killing vectors.

Assuming the existence of two Killing vectors $\partial_4$ (timelike) and
$\partial_5$ (spacelike), the $GL(2,R)$-covariant five-to-three reduction
is achieved by
\begin{equation}\label{st5}
ds^2 = \lambda_{ab}(dx^a + a_i^adx^i)(dx^b
+ a_j^bdx^j) + \tau^{-1}h_{ij}dx^idx^j\,,
\end{equation}
where $a,b=4,5$, $i,j=1,2,3$, and $\tau \equiv -$det$\lambda$. The reduction
of the five-dimensional Einstein equations follows the same path as in the
case of E4, the duality equation (\ref{duale4}) generalizing to
\begin{equation}\label{duale5}
\tau\lambda_{ab}\nabla\wedge\vec{a}^b = \nabla\omega_a\,.
\end{equation}
The result is a gravitating sigma model for a five-dimensional
target space (three ``gravielectric''  potentials $\lambda_{ab}$ and
two ``gravimagnetic'' potentials $\omega_a$). Maison determined the
eight Killing vectors of this target space, and identified the
symmetry group as $SL(3,R)$. He then showed that this group acts
bilinearly on the symmetric, unimodular matrix potential
\begin{equation}\label{maimat}
\chi = \left(\begin{array}{cc}
\lambda - \tau^{-1}\omega\omega^T &
\tau^{-1}\omega \\ \tau^{-1}\omega^T & -\tau^{-1}
\end{array}\right)\,,
\end{equation}
where $\lambda$ is a $2\times2$ block, and $\omega$ a 2-component column
matrix.

In terms of $\chi$, the reduced field equations
\begin{eqnarray}
\nabla(\chi^{-1}\nabla\chi) &=& 0\,, \label{divJ5} \\
R_{ij}(h) &=& \frac14{\rm Tr}(\chi^{-1}\partial_i\chi
\chi^{-1}\partial_j\chi)\,, \label{Rij5}
\end{eqnarray}
are manifestly invariant under $G = SL(3,R)$. Assuming the five-dimensional
metric (\ref{st5}) to be asymptotically Minkowskian with signature $-++++$,
the Maison matrix (\ref{maimat}) goes asymptotically to the constant matrix
\begin{equation}\label{etabs}
\chi_{\infty} = \eta_{BS} = \left(\begin{array}{ccc}
-1 & 0 & 0 \\ 0 & 1 & 0 \\ 0 & 0 & -1
\end{array}\right)\,.
\end{equation}
At this point we should stress that our seemingly natural assumption
of an asymptotically Minkowskian five-metric implies $g_{55}(\infty)
=$ constant. This is indeed the case for five-dimensional black
strings (BS), but not for five-dimensional black holes (BH), in
which case the fifth coordinate is an angle on the three-sphere; we
shall return to this question in subsection \ref{subgiusax}. The
asymptotic behavior (\ref{etabs}) is preserved by the isotropy
subgroup at infinity $H = SO(2,1)$. Thus, the five-dimensional
target space for E5 is $SL(3,R)/SO(2,1)$.

\subsection{First applications}
The first application of this formalism was given in 1982 by
Dobiasch and Maison \cite{DM} , who obtained all the 2-stationary
spherically symmetric solutions of E5 by direct solution of the matrix
differential equations (\ref{divJ5}) and (\ref{Rij5}), generalizing
previous static spherically symmetric solutions given by Leutwyler
\cite{leut} and Chodos and Detweiler \cite{CD82}.

In 1986 I proposed a scheme \cite{statio} to relate stationary
solutions of five-dimensional general relativity (invariance group
$SL(3,R)$) and stationary solutions of four-dimensional
Einstein-Maxwell theory (invariance group $SU(2,1)$). Stationary
solutions of EM4 depending on at most two real potentials belong to sectors
invariant under subgroups of $SU(2,1)$. These are locally isomorphic
to subgroups of $SL(3,R)$:
\begin{eqnarray*}
\underline{SU(2,1)} & & \underline{SL(3,R)} \\
&& \\
SU(1,1) & \qquad \approx \qquad & SL(2,R) \\
SO(2,1) & \qquad = \qquad & SO(2,1) \\
SU(2) & \qquad \approx \qquad & SO(3) \\
U(1) & \qquad \approx \qquad & O(2)\,,
\end{eqnarray*}
implying a correspondence between stationary solutions of the two
theories. In the axisymmetric case, this correspondence could be
made precise, leading to theorems relating exact solutions of EM4
and E5. Several of these relations were already known, however the
investigation of the case $SU(2) \approx SO(3)$ led to a new class
of 2-stationary five-dimensional spacetimes generated from the class
${\cal E} = +1$ of solutions of the Einstein-Maxwell equations. This
construction was applied to generate from the massless Kerr-Newman
solution a three-parameter family of geodesically complete,
asymptotically flat solutions of E5, rotating generalizations of
static axisymmetric Lorentzian wormhole solutions constructed from
the Chodos-Detweiler wormhole \cite{CD82} in \cite{axi}.

\subsection{Multicenter solutions}\label{sub33}
Let me first introduce the basic procedure \cite{spat,bps} for a generic
gravitating sigma model, before describing briefly its application to E5.
Consider the case of solutions depending on a single real potential,
$M = M[\sigma(x)]$. As shown in \cite{neukra}, this potential
can be chosen to be harmonic,
\begin{equation}\label{harm}
\nabla^2\sigma = 0\,.
\end{equation}
Then the equations (\ref{divJ}) and (\ref{Rij}) reduce to
\begin{eqnarray}
\frac{d}{d\sigma}\left(M^{-1}\frac{dM}{d\sigma}\right)&=&0\,, \\
R_{ij} &=& \frac14{\rm Tr}\left(M^{-1}\frac{dM}{d\sigma}\right)^2
\partial_i\sigma\partial_j\sigma\,.
\end{eqnarray}
The first of these equations is the geodesic equation for the target space
metric (\ref{tarmet}) with $\sigma$ the affine parameter. It is solved by
\begin{equation}\label{etaA}
M = \eta{\rm e}^{A\sigma}\,,
\end{equation}
where $\eta \in G/H$ and $A \in {\rm Lie}(G)$ are constant matrices. If
$\sigma(\infty)=0$, then $\eta = M_{\infty}$. In terms of the solution
(\ref{etaA}), the target space metric (\ref{tarmet}) and
the three-dimensional Einstein equations (\ref{Rij}) become
\begin{equation}
dl^2=\frac{1}{4}{\rm Tr}(A^2)\,d\sigma^2\,, \qquad
R_{ij}=\frac{1}{4}{\rm Tr}(A^2)\partial_i\sigma\partial_j\sigma\,,
\end{equation}
showing that the sign of the spatial curvature, hence the nature of
the three-geometry, depends on the sign of the constant ${\rm
Tr}(A^2)$. Black holes necessarily belong to the class of spacelike
target space geodesics (${\rm Tr}(A^2) > 0$), while Lorentzian
wormholes \cite{worm} necessarily belong to the class of timelike
target space geodesics (${\rm Tr}(A^2) < 0$). Null target space
geodesics,
\begin{equation}\label{anti}
{\rm Tr}(A^2) = 0\,,
\end{equation}
lead to solutions with a flat reduced three-space. The harmonic condition
(\ref{harm}) reduces in this case to the linear Laplace equation, the
solutions of which can be linearly superposed, yielding multicenter solutions
\begin{equation}
\sigma(\vec{x}) = \sum_{\alpha}\frac{c_{\alpha}}{|\vec{x}-\vec{a_{\alpha}}|}
\,.
\end{equation}

In the case of E5, the symmetry and unimodularity of the Maison matrix imply
the conditions
\begin{equation}\label{condA}
A^T = \eta A\eta\,, \quad {\rm Tr}A = 0\,.
\end{equation}
With $\eta$ given by (\ref{etabs}), the generic matrix $A$ may be
parametrized as
\begin{equation}\label{Agen}
A = 2\left(\begin{array}{ccc}
-M-\Sigma/\sqrt3 & -Q & N \\
Q & 2\Sigma/\sqrt3 & P \\
N & -P & M - \Sigma/\sqrt3
\end{array}\right)\,,
\end{equation}
where the parameters $M$, $N$, $\Sigma$, $Q$ and $P$ are proportional to
the mass, NUT charge, scalar charge, electric charge, and magnetic charge.
The constraint (\ref{anti}) translates into
\begin{equation}
M^2 + N^2 +\Sigma^2 -Q^2 -P^2 = 0\,,
\end{equation}
which generalizes the antigravity condition of Sherk \cite{sherk,gibbons82}
expressing the balance between ``scalar'' (attractive) forces and ``vector''
(repulsive) forces.

The matrix $A$ is a solution of its characteristic equation which, owing to
the constraints (\ref{anti}) and (\ref{condA}), reduces to
\begin{equation}
A^3 = {\rm det}A\,.
\end{equation}
As discussed in \cite{spat}, this leads to three classes of null target space
geodesics according to the rank $r(A)$ of $A$:
\begin{enumerate}
\item\label{r3} r(A) = 3 ($A^3 = 1$). This class contains, among others,
regular static dyon solutions.
\item\label{r2} r(A) = 2 ($A^3 = 0$, $A^2 \neq 0$). This class contains
static dyons with equal electric and magnetic charges and vanishing scalar
charge sitting in a four-dimensional geometry which is (multiple) extreme
Reissner-Nordstr\"om.
\item\label{r1} r(A) = 1 ($A^2 = 0$). This class contains systems of
electric or magnetic \cite{mono} monopoles.
\end{enumerate}

The construction (\ref{etaA}) may be generalized to the case of several
harmonic functions \cite{bps}. In the case of E5 \cite{spat}, multicenter
solutions depending on two real harmonic potentials $\sigma$ and $\phi$
are totally geodesic surfaces of the target space
\begin{equation}\label{etaAB}
\chi = \eta{\rm e}^{A\sigma}{\rm e}^{A^2\phi}\,,
\end{equation}
where the matrix $A$ is of rank 2 (class \ref{r2}). Appropriate choices of
this matrix lead, for harmonic functions which are the real and imaginary
part of the complex potential
\begin{equation}
V(\vec{x}) = \sum_{\alpha}\frac{c_{\alpha}}{|\vec{x}-\vec{a_{\alpha}}
-i\vec{b_{\alpha}}|}\,,
\end{equation}
to solutions describing systems of rotating electric or magnetic monopoles,
or of rotating dyons, generalizing the classes \ref{r1} and \ref{r2} above.
Multiple cosmic string solutions of the five-dimensional Einstein-Gauss-Bonnet
equations \cite{love} were also constructed from the ansatz (\ref{etaAB}) in
\cite{GB}.

\subsection{Rotating dyonic black strings}
Stationary black strings in five dimensions (or Kaluza-Klein black holes)
were investigated in 1995 by Rasheed \cite{rasheed}. The Maison matrix for
asymptotically flat static black strings with regular horizons \cite{GW}
is of the form (\ref{etaA}) with ${\rm det}A = 0$. This can be obtained
from the Schwarzschild Maison matrix, such that $N=\Sigma=Q=P=0$ in
(\ref{Agen}) and
\begin{equation}
\sigma = -\frac1{2M}\ln\left(1-\frac{2M}r\right)\,,
\end{equation}
by an $SO(2,1)$ transformation. Similarly, rotating dyonic black strings
may be generated from the Kerr metric by the global group transform
\begin{equation}
\chi = P^T \chi_K P\,,
\end{equation}
where $\chi_K$ is the Maison matrix for the Kerr black string
\begin{equation}\label{bsk}
ds_{(5)}^2 = (dx^5)^2 + ds_K^2\,,
\end{equation}
with $ds_K^2$ the four-dimensional Kerr metric, and $P$ an $SO(2,1)$
matrix. Rasheed actually used only the subclass of transformations $P$
constrained so that the final NUT charge $N$ vanishes, yielding a
four-parameter (mass, angular momentum, and electric and magnetic charges)
family of rotating black strings.

\subsection{Relating black strings and black holes}
Besides black strings, with a topologically $R \times S^2$ horizon,
five-dimensional general relativity also admits black hole solutions
\cite{mype}, with the horizon topology $S^3$. The static spherically
symmetric black hole is given by the Tangherlini solution \cite{tang},
\begin{equation}\label{tan1}
ds_T^2 = -\bigg(1-\frac{\mu}{\rho^2}\bigg)dt^2 +
\bigg(1-\frac{\mu}{\rho^2}\bigg)^{-1}d\rho^2 +
\rho^2d\Omega_3^2\,,
\end{equation}
where
\begin{equation}
 d\Omega_3^2 =
\frac14\bigg[(d\eta-\cos\theta d\varphi)^2 + d\theta^2 +
\sin^2\theta d\varphi^2\bigg]
\end{equation}
is the three-sphere line element. Putting  $\rho^2 = 4mr$, $\eta = x^5/m$,
with $m^2 = \mu/8$, transforms the line element (\ref{tan1}) to
\begin{equation}\label{tan2}
ds_T^2 = -\frac{r-2m}{r}\,dt^2 +
\frac{r}{m}(dx^5-m\cos\theta d\varphi)^2 + \frac{m}{r-2m}\bigg[dr^2 +
r(r-2m)\bigg(d\theta^2 + \sin^2\theta d\varphi^2\bigg)\bigg]\,.
\end{equation}
As observed in \cite{newdil2}, this has the same reduced three-dimensional
metric as the four-dimensional Schwarzschild black string,
\begin{equation}\label{sch}
ds_S^2 = -\frac{r-2m}{r}\,dt^2 +
(dx^5)^2 + \frac{r}{r-2m}\bigg[dr^2 + r(r-2m)\bigg(d\theta^2 +
\sin^2\theta d\varphi^2\bigg)\bigg]\,,
\end{equation}
so the two corresponding Maison matrices must be related by an $SL(3,R)$
transformation
\begin{equation}\label{transfST}
\chi_T = P_{ST}^T\chi_SP_{ST}\,.
\end{equation}
The transformation matrix $P_{ST}$, which was determined in
\cite{newdil2}, does not belong to the subgroup $SO(2,1)$ because
the metrics (\ref{sch}) and (\ref{tan2}) have different asymptotic
behaviors, so that the matrix $\chi_T$ goes for $r \to \infty$ to a
constant matrix $\eta_{BH}$ (given in the next subsection) different
from the matrix $\eta_{BS}$ of (\ref{etabs}).

As all the black string metrics have the same asymptotic behavior as
(\ref{sch}), and all the black hole metrics have the same asymptotic
behavior as (\ref{tan2}), one expects that the transformation
(\ref{transfST}) will more generally transform black strings into
black holes:
\begin{equation}\label{bsbh}
\chi_{BH} = P_{ST}^T\chi_{BS}P_{ST}\,.
\end{equation}
Indeed, it was found in \cite{newdil2} that the action of this
transformation on the Kerr black string (\ref{bsk}) led to the
Myers-Perry black hole with opposite angular momenta\footnote{To 
conform with the conventions of \cite{giusax}, we use here angular 
momentum parameters related to those (primed) of \cite{newdil2} by $a_{\pm}
= \mp a_{\mp}'$.}, $a_+ = -a_-$. It was also observed in
\cite{newdil2} (Sect. 6) that the reduced three-dimensional metric
of the generic Myers-Perry black hole with arbitrary $a_+$ and $a_-$
again coincided with that of the Kerr black string. In the
special case of the Myers-Perry black hole with {\em equal} angular
momenta, $a_+ = a_-$, the reduced four-dimensional
metric was static (dyonic), and the reduced three-dimensional metric
was the same as that of the Schwarzschild black string. Unfortunately, 
the full significance of this observation was missed (see next section).

To be complete, let me mention an alternate black string-black hole
correspondence which was also given in \cite{newdil2}. This proceeds
via the standard five-to-four Kaluza-Klein reduction to $\alpha^2 =
3$ Einstein-Maxwell-dilaton theory, according to the diagram
$$
\begin{array}{ccc}
\mbox{\rm 5D static BH} & & \mbox{\rm 5D static BS} \\
\mbox{\rm (Tangherlini)} & & \mbox{\rm (Schwarzschild$^{\prime}$)} \\
\downarrow & & \uparrow \\
\mbox{\rm 4D NAF static} & & \mbox{\rm 4D NAF
static} \\
\mbox{\rm magnetic BH} & \longrightarrow & \mbox{\rm electric BH}
\end{array}
$$
where the downarrow (uparrow) stands for Kaluza-Klein reduction
(oxidation), and the rightarrow stands for the four-dimensional
electric-magnetic duality relating the non-asymptotically flat (NAF)
magnetic and electric black holes. The Schwarzschild$^{\prime}$ metric is
$$ ds_S^{'2} = -2\frac{r-2m}{r}(dt-\frac12dx^5)^2 + \frac12(dx^5)^2 +
\frac{r}{r-2m}[dr^2 + r(r-2m)(d\theta^2 + \sin^2\theta
d\varphi^2)]\,. $$ This correspondence was extended in
\cite{newdil2,cedric} to relate Myers-Perry black holes
with two angular momenta and Rasheed dyonic black strings with NUT
charge.

\subsection{Relating static and rotating solutions}\label{subgiusax}
Giusto and Saxena independently observed in \cite{giusax} that the asymptotic
Maison matrix is different for black strings and for black holes.
For the Tangherlini metric (\ref{tan2}), $\omega_4=0$, but
$\omega_5=r/m + b$ (where $b$ is some additive constant) diverges at
infinity with $\tau^{-1}\omega_5 \to 1$. The Maison matrix $\chi_T$ 
goes to a finite non-diagonal
limit $\eta_{BH}$ at infinity, and the constant $b$ may be chosen so
that $\chi_{55}(\infty) = 0$, leading to
\begin{equation}\label{etabh}
\eta_{BH} = \left(\begin{array}{ccc} -1 & 0 & 0 \\ 0 & 0 & 1
\\ 0 & 1 & 0
\end{array}\right)\,.
\end{equation}
The two matrices $\eta_{BH}$ and $\eta_{BS}$ are of course related
by the general black string/black hole transformation
(\ref{transfST})
\begin{equation}
\eta_{BH} = P_{ST}^T\eta_{BS}P_{ST}\,,
\end{equation}
with the transformation matrix (for the choice of the additive
constant $b$ just mentioned)
\begin{equation}\label{P3ST}
P_{ST} = \left(\begin{array}{ccc} 1 & 0 & 0 \\ 0 & 1/\sqrt2 &
1/\sqrt2 \\ 0 & -1/\sqrt2 & 1/\sqrt2
\end{array}\right)\,.
\end{equation}

The isotropy group $SO(2,1)$ preserving $\eta_{BH}$ is generated by
three transformations $M_{\alpha}$, $M_{\beta}$ and $M_{\gamma}$,
different from the Rasheed transformations preserving $\eta_{BS}$.
While the transformations $M_{\beta}$ and $M_{\gamma}$ are trivial
(generalized gauge transformations), Giusto and Saxena showed that
the transformation $M_{\alpha}$ could be used to generate the
Myers-Perry black hole from the static Tangherlini black hole in
three steps:
\begin{enumerate}
\item Act with $M_{\alpha}$ on the Tangherlini metric (\ref{tan1})
(written in a form which exhibits the symmetry between the $S^3$
angles $\eta$ and $\varphi$)
\begin{equation}\label{tan3}
ds_T^2 = -\bigg(1-\frac{\mu}{\rho^2}\bigg)dt^2 +
\bigg(1-\frac{\mu}{\rho^2}\bigg)^{-1}d\rho^2 + \frac{\rho^2}4
(d\theta^2 + d\eta^2 + d\varphi^2 - 2\cos\theta d\eta d\varphi)
\end{equation}
reduced with respect to the Killing vectors $\partial_t$ and
$\partial_{\eta}$. This leads to the Myers-Perry metric with
equal angular momenta, $a_+ = a_-$.
\item "Flip" the angles $\eta\leftrightarrow \varphi$. This amounts
to reducing the Myers-Perry metric with $a_+ = a_-$ with respect to
the Killing vectors $\partial_t$ and $\partial_{\varphi}$ instead of
$\partial_t$ and $\partial_{\eta}$ (a finite Geroch
transformation!) or, equivalently, to replacing the Myers-Perry
metric with equal angular momenta by the Myers-Perry metric with
opposite angular momenta.
\item Act on this with a second transformation $M_{\alpha'}$,
leading to a generic Myers-Perry metric with $a_+ \neq \pm a_-$.
\end{enumerate}

More generally, as discussed in \cite{giusax}, the combined
transformation $M_{\alpha'} \ast {\rm Flip} \ast M_{\alpha}$ can be
used to generate a 2-rotating solution from any given seed static
solution. An interesting (but perhaps technically involved) exercize
would be take as a seed the static (singular) black ring of Emparan
and Reall \cite{ER02a} with $S^1\times S^2$ horizon, and generate
{\em \`a la} Giusto-Saxena:
\begin{itemize}
\item the black ring rotating along $S^1$ \cite{ER02b};
\item the black ring rotating along $S^2$ \cite{MIF};
\item the black ring with two angular momenta \cite{pomsen}.
\end{itemize}

\subsection{Summary}
We have seen that E5 reduced to three dimensions leads to the 
$SL(3,R)/SO(2,1)$ gravitating sigma model. The two main applications 
of this sigma model are:
\begin{enumerate}
\item The generation by $SL(3,R)$ transformations of rotating dyonic black 
strings from the Kerr black string (Rasheed), black holes from black strings
(Cl\'ement-Leygnac), and rotating black holes from static black holes 
(Giusto-Saxena). So, all rotating five-dimensional black strings and black 
holes can be generated from the four-dimensional Schwarzschild solution.
\item The construction of rotating multicenter solutions from null 
totally geodesic surfaces of the target space.
\end{enumerate}

\section{The case of minimal five-dimensional supergravity (EM5)}
\label{sec4}
\setcounter{equation}{0}

\subsection{The $G_{2(2)}$-based sigma model}
The bosonic sector of five-dimensional minimal supergravity 
\cite{cremmer81,CN} is defined
by the Einstein-Maxwell-Chern-Simons action (see also Jutta Kunz's talk at
this Seminar)
\begin{equation}\label{em5} 
S_5 = -\frac1{16\pi G_5}\int d^5x \bigg[\sqrt{|g|} \bigg(R +
\frac14F^{\mu\nu}F_{\mu\nu}\bigg) +
\frac1{12\sqrt3}\epsilon^{\mu\nu\rho\sigma\lambda}F_{\mu\nu}
F_{\rho\sigma}A_{\lambda}\bigg]\,, 
\end{equation} 
where $F = dA$, and $\epsilon^{\mu\nu\rho\sigma\lambda}$ is 
the five-dimensional antisymmetric  symbol. It has been known for 
some time \cite{mizoh,CJLP,GNPP} that reduction of (\ref{em5}) to three
Euclidean dimensions leads to the $G_{2(2)}/SL(2,R) \times SL(2,R)$
coset. I will here outline the five-to-three reduction, the identification 
of the coset and the construction of coset representatives following
the approach of \cite{5to3}.
\subsubsection{Five-to-three reduction.}
The five-dimensional metric fields of (\ref{em5}) are broken down by the 
Kaluza-Klein ansatz (\ref{st5}) to the scalars $\lambda_{ab}$ and the 
Kaluza-Klein vectors $a_i^a$, and the five-dimensional electromagnetic 
potential is broken down according to  
\begin{equation}  
A_{(5)} = \sqrt3(\psi_a dx^a + A_idx^i)\,.  
\end{equation}
The space components of the Maxwell-Chern-Simons equations allow the 
dualization of the vector potentials $A_i$ to the magnetic scalar potential
$\mu$,
\begin{equation}
\tau\left(\nabla\wedge\vec{A} + \vec{a}^a\wedge\nabla \psi_a\right)
- \epsilon^{ab}\psi_a\nabla\psi_b  = \nabla\mu\,,
\end{equation}
while the mixed ($ai$) Einstein equations now lead to the duality equations 
for the gravimagnetic scalar potentials $\omega_a$
\begin{equation}
\tau\lambda_{ab}\nabla\wedge\vec{a}^b + \psi_a(3\nabla\mu 
+ \epsilon^{bc}\psi_b\nabla\psi_c) = \nabla\omega_a\,.
\end{equation}
The remaining field equations are those of a gravitating sigma model, with 
the eight-dimensional target space $\cal T$ of metric:
\begin{equation}\label{tarem5} 
dS^2 = \frac12
Tr(\lambda^{-1}d\lambda\lambda^{-1}d\lambda) + \frac12\tau^{-2}d\tau^2
- \tau^{-1}V^T\lambda^{-1}V +
3\left(d\psi^T\lambda^{-1}d\psi - \tau^{-1}\eta^2\right) \,, 
\end{equation} 
where
\begin{equation} 
\eta = d\mu +   \epsilon^{ab} \psi_a d\psi_b \,, \qquad V_{a} =
d\omega_a - \psi_a\left(3d\mu + \epsilon^{bc}\psi_bd\psi_c\right)\,.
\end{equation}
\subsubsection{Isometries of ${\cal T}$.}
This metric admits nine manifest Killing vectors, grouped into the $GL(2,R)$
multiplets:
\begin{itemize}
\item a quadruplet (generators of $gl(2,R)$ transformations in the ($x^4,x^5$) 
plane)
\begin{equation} 
{M_a}^b =
2\lambda_{ac}\frac{\partial}{\partial\lambda_{cb}} +
\omega_a\frac{\partial}{\partial\omega_{b}} +
\delta_a^b\omega_c\frac{\partial}{\partial\omega_{c}} +
\psi_a\frac{\partial}{\partial\psi_{b}} +
\delta_a^b\mu\frac{\partial}{\partial\mu}\,, 
\end{equation}

\item a doublet and a singlet (translations of the ``magnetic'' coordinates)
\begin{equation} 
N^a = \frac{\partial}{\partial\omega_a}\,, \qquad Q =
\frac{\partial}{\partial\mu}\,,
\end{equation}

\item and a doublet (gauge transformations of the ``electric'' coordinates 
$\psi_a$) 
\begin{equation} 
R^a = \frac{\partial}{\partial\psi_a} +
3\mu\frac{\partial}{\partial\omega_a} -
\epsilon^{ab}\psi_b\left(\frac{\partial}{\partial\mu} +
\psi_c\frac{\partial}{\partial\omega_c}\right) \,.
\end{equation}   

\end{itemize}

This is enough information to determine the full isometry group $G$ 
\cite{5to3}, provided one makes the two assumptions:
\begin{enumerate}
\item $G$ contains the subgroup $SL(3,R)$ (this is motivated by the fact that 
EM5 can be consistently truncated to E5 by taking $A_{(5)}=0$);
\item Lie($G$) is minimal.
\end{enumerate}
These, together with the Jacobi identities, lead to the 
conclusion\footnote{The question of the symmetry group for a value of the 
Chern-Simons coupling constant different from the supergravity value 
\cite{kunz} in (\ref{em5}) remains open.} that the 
algebra is $g_2$ \cite{gilmore}. This has 14 generators, the nine 
manifest Killing 
vectors given above, together with the five hidden Killing vectors $L_a$, 
$P_a$, and $T$. These are determined by solving the Lie brackets, up to 
a single integration constant, which is fixed by enforcing that $T$ is a 
Killing vector of (\ref{tarem5}). The result is  
\begin{eqnarray} 
T &=& \left[2\mu\lambda_{bc} +
6\epsilon^{de}\lambda_{bd}\psi_c\psi_e\right]
\frac{\partial}{\partial\lambda_{bc}} +
\left[3\mu\omega_{b} + 3\tau\psi_b -
\epsilon^{cd}\omega_{c}\psi_b\psi_d +
4\tau\lambda^{cd}\psi_b\psi_c\psi_d\right]
\frac{\partial}{\partial\omega_{b}} \nonumber \\ &+& \left[\omega_b +
\mu\psi_{b} + 2\epsilon^{cd}\lambda_{bd}\psi_c\right]
\frac{\partial}{\partial\psi_{b}} + \left[\mu^2 +
\tau - \epsilon^{bc}\omega_{b}\psi_c +
2\tau\lambda^{bc}\psi_b\psi_c\right]
\frac{\partial}{\partial\mu}\,.
\end{eqnarray}
The remaining hidden Killing vectors can be determined from $\left[R^a,T\right]
= 2\epsilon^{ab}P_b$ and $\left[P_a,T\right] = 3L_a$.

Taking into account the signature of the target space metric (\ref{tarem5}), 
the isometry group is the real noncompact form $G_{2(2)}$ of the exceptional 
group $G_2$. The root diagram of the $g_2$ algebra is shown in Fig.\ 1. The 
generators of the Cartan subalgebra are
\begin{equation}
H_1 = ({M_4}^4 + {M_5}^5)/\sqrt6\,, \quad H_2 = ({M_4}^4 - {M_5}^5)/\sqrt2\,.
\end{equation}
The $sl(2,R)$ subalgebra contains these together with the six outermost roots.
\begin{figure}
\centering
\includegraphics[height=8cm]{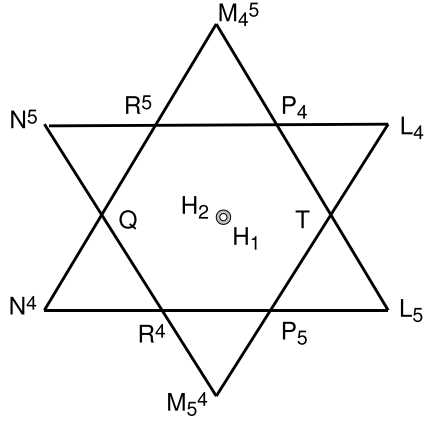}
\caption{Root diagram of $g_2$}
\label{fig:1}      
\end{figure}
\subsubsection{Coset representative.}
The real form of $g_2$ may be represented in terms of real $7\times7$
matrices obtained from the $Z$ matrices of \cite{GG} by omitting $i$'s.
From the structure of the $gl(2,R)$ generators, one infers that the coset
matrix representative in the vacuum ($\psi = \mu = 0$) sector is a real 
$7\times7$ matrix of the block form
\begin{equation}\label{M1} 
{\cal M}_1 = \left(\begin{array}{ccc}
\chi & 0 & 0 \\ 0 & \chi^{-1} & 0 \\  0 & 0 & 1 \end{array}\right)\,,
\end{equation}
where $\chi$ is the Maison matrix (\ref{maimat}). The coset representative 
for the full target space may be obtained \cite{g2,5to3} by the local group 
transformation
\begin{equation}  
{\cal M} = N^T{\cal M}_0N 
\end{equation}
acting on the static ($\omega = 0$) matrix
\begin{equation}  
{\cal M}_0 = \left(\begin{array}{ccccc} 
\lambda & 0 & 0 & 0 & 0 \\ 0
& -\tau^{-1} & 0 & 0 & 0 \\ 0 & 0 & \lambda^{-1} & 0 & 0 \\ 0 & 0 & 0
& -\tau & 0 \\ 0 & 0 & 0 & 0 & 1
\end{array}\right)
\end{equation}
with
\begin{equation} 
N = {\rm e}^{\psi_ar^a}{\rm e}^{\mu q}{\rm e}^{\omega_an^a}\,,
\end{equation}
where $n^a$, $q$, $r^a$ are the matrix representatives of the manifest Killing
vectors $N^a$, $Q$, $R^a$.
The resulting matrix representative\footnote{Another, equivalent $7\times7$
matrix representation of the $G_{2(2)}/SL(2,R)^2$ coset was independently 
given in \cite{nikolai}.} has the  symmetrical block structure 
\begin{equation} 
{\cal M} = \left(\begin{array}{ccc} 
A & B & \sqrt2U \\ B^T & C & \sqrt2V \\
\sqrt2U^T & \sqrt2V^T & S
\end{array}\right)\,,
\end{equation}
where $A$ and $C$ are symmetrical $3\times3$ matrices, $B$ is a
$3\times3$ matrix, $U$ and $V$ are 3-component column matrices, and
$S$ a scalar. The inverse matrix is given by
\begin{equation} 
{\cal M}^{-1} = K{\cal M}K\,, \qquad
K = \left(\begin{array}{ccc}0&1&0\\1&0&0\\0&0&-1\end{array}\right)\,.  
\end{equation}

\subsection{Applications}
The first, obvious application \cite{g2} was to generate charged solutions 
of EM5 from seed vacuum solutions by the global group transform
\begin{equation}\label{transfC}
{\cal M}' = C^T{\cal M}C\,,
\end{equation}
with
\begin{equation}\label{C}
C = {\rm e}^{\beta(p_4+r^4)}
\end{equation}
the charge-generating transformation. This was tested on the example of a
Myers-Perry vacuum seed \cite{mype}, yielding the charged rotating black 
hole first given in \cite{CY96} and rederived, in a different parametrization, 
in \cite{chong05}. The action of the transformation (\ref{transfC}) on the 
neutral black ring with two angular momenta \cite{pomsen} led to a doubly 
rotating charged black ring. However the resulting five-dimensional solution
($g'$, $A'$) was singular, due to the presence of Dirac-Misner strings. The 
same transformation was also applied recently to the generation of rotating 
black strings with Maxwell electric charge from Rasheed black strings with 
vanishing Kaluza-Klein electric charge \cite{TYM}. An alternate route to the
generation of such charged black strings would be to apply the inverse black 
string/black hole transformation (\ref{bsbh}), extended to EM5 (see subsection 
\ref{subkerr}), to charged black holes.

It has recently been pointed out that there is a shorter route to charge 
generation in
EM5. Minimal five-dimensional supergravity reduced to four dimensions has a 
global $SL(2,R)$ symmetry, and a certain $O(1,1) \in SL(2,R)$ transformation
does the same job as the charging transformation $C$, without all the 
cumbersome $G_{2(2)}$ machinery \cite{LMP08}.   

The potentialities of the $G_{2(2)}$ generating technique could be more fully
exploited by extending the Giusto-Saxena approach to this case. The asymptotic
limit $\eta_{BH}$ of the matrix $\cal M$ for black holes is (\ref{M1}) with 
$\chi$ given by (\ref{etabh}). The isotropy subgroup $SL(2,R)\times SL(2,R)$ 
preserving $\eta_{BH}$ \cite{spinem5} now contains three trivial (gauge)
transformations, and the three non-trivial transformations:
\begin{eqnarray}
S &=& {\rm e}^{\alpha(l_4+{m_5}^4)}\,, \nonumber \\
C &=& {\rm e}^{\beta(p_4+r^4)}\,, \\
D &=& {\rm e}^{\gamma(p_5-t)}\,. \nonumber 
\end{eqnarray}
$S$ is the original Giusto-Saxena spin-generating transformation, $C$ is 
our electric charge-generating transformation (\ref{C}), and the 
transformation $D$ generates a dipole charge. One could in principle 
combine these three transformations together with flips to generate a 
five-parameter black ring from the uncharged static black ring of \cite{ER02a}.
This solution-generating technique could also be applied to the recently
discovered black lens solution \cite{chenteo}.

\subsection{Generating rotating solutions via 5D Bertotti-Robinson}
\label{subkerr}

\subsubsection{Generating rotating solutions in EM4}
Let me first recall the sigma-model technique to generate rotating solutions
outlined in \cite{kerr}. As recalled in \ref{sub21}, four-dimensional 
Einstein-Maxwell theory reduced (with respect to $\partial_t$) to three
dimensions leads to the $SU(2,1)/S[U(2)\times U(1)]$ sigma model. According 
to Geroch \cite{geroch}, in the stationary axisymmetric case one can in 
principle combine finite global $SU(2,1)$ transformations with finite linear 
transformations in the plane of the two Killing vectors $\partial_t$ and 
$\partial_{\varphi}$ to generate new solutions. The problem is that the only 
such linear transformation which does not lead to the appearance of closed 
timelike curves and/or conical singularities is the transformation 
${\cal R}(\Omega,\gamma)$ (transition to a uniformly rotating frame combined 
with a time dilation),
\begin{eqnarray}\label{R}
d\varphi' &=&  d\varphi - \Omega\,dt\,,   \nonumber \\
dt' &=&  \gamma^{-1}\,dt\,.
\end{eqnarray} 
which, in the case of an asymptotically Minkowskian seed solution,
drastically changes the asymptotic behavior (appearance of a ``centrifugal 
force''). The solution to this problem, 
as given in \cite{kerr}, is to combine the transformation (\ref{R}) with 
an $SU(2,1)$ transformation $\Pi$ also changing the asymptotic behavior.

Consider the Bertotti-Robinson solution to EM4 \cite{br}, describing the
$AdS_2\times S^2$ spacetime generated by a constant electric field,
\begin{eqnarray}\label{bert}
d\hat{s}^2 &=& \left(1-\frac{x^2}{m^2}\right)\,dt^2 +
\left(\frac{x^2}{m^2}-1\right)^{-1}\bigg[dx^2 + (x^2-m^2)\bigg(d\theta^2 +
\sin^2\theta d\varphi^2\bigg)\bigg]\,, \nonumber \\
\hat{A} &=& -\frac{x}{m}\,dt\,. 
\end{eqnarray}
The reduced three-dimensional metric in (\ref{bert}) is the same as that of 
the Schwarzschild metric ($x = r-m$) 
\begin{equation}
ds_S^2 = -\frac{x-m}{x+m}\,dt^2 +
\frac{x+m}{x-m}\bigg[dr^2 + (x^2-m^2)\bigg(d\theta^2 +
\sin^2\theta d\varphi^2\bigg)\bigg]\,,
\end{equation}
so the two solutions are related by a transformation $\Pi \in SU(2,1)$, given 
in \cite{kerr}. The global coordinate transformation ${\cal R}(\Omega,\gamma)$ 
acting on the Bertotti-Robinson solution now leads to a solution 
($d\hat{s}^{'2}$, $\hat{A}'$) (the same Bertotti-Robinson spacetime and 
electromagnetic field described in a different coordinate system) with the 
same asymptotic behavior, for instance,
\begin{equation}
\hat{g}_{tt}' = \gamma^2(1+m^2\Omega^2\sin^2\theta-r^2/m^2)\,.
\end{equation}
One can now return to the asymptotically flat world by the action on this 
primed Bertotti-Robinson solution of the inverse transformation $\Pi^{-1}$, 
leading (for the choice\footnote{For a generic value of $\gamma$ one obtains 
the Kerr-Newman solution.} $\gamma = (1+m^2\Omega^2)^{-1}$) to the Kerr 
solution:
$$
\begin{array}{cccc}
& & {\cal R}(\Omega,\gamma) & \\
({\rm as.}\; AdS_2\times S^2)\;\; & \mbox{\rm Bertotti-Robinson} 
& \longrightarrow & \mbox{\rm Bertotti-Robinson$'$} \\
& \Pi\,\uparrow & & \Pi^{-1}\,\downarrow \\
({\rm as.}\; M_4) & \mbox{\rm Schwarzschild} & & \mbox{\rm Kerr}
\end{array}
$$
More generally, the combined transformation $\Sigma(\Omega,\gamma)=
\Pi^{-1}\ast {\cal R}(\Omega,\gamma)\ast \Pi$ acting on a static 
asymptotically flat solution leads to a rotating asymptotically flat 
solution (examples are given in \cite{kerr}).  

\subsubsection{Generalization to EM5}
Reduction of EM5 to four spacetime dimensions leads to an Einstein theory 
with two coupled abelian gauge fields, the reduced Maxwell field $F$ and 
the Kaluza-Klein field $G$, a dilaton $\phi$ and an axion $\kappa$. This 
theory can be consistently truncated to four-dimensional Einstein-Maxwell 
theory (EM4) by enforcing the constraints
\begin{equation}\label{em4}
\phi = 0\,, \quad \kappa = 0\,, \quad G = \frac1{\sqrt3}\star F\,.
\end{equation}
After further reduction to three dimensions, one finds \cite{spinem5} that
these constraints are preserved by eight infinitesimal transformations which 
generate the Lie algebra of $SU(2,1)$ (the isometry group of EM4 reduced to 
3D). 

Conversely, any solution of EM4 can be lifted to a solution of EM5 satisfying
the constraints (\ref{em4}). Applying this lifting procedure to the 
four-dimensional Bertotti-Robinson solution (\ref{bert}) ($AdS_2\times S^2$), 
we obtain \cite{spinem5} the five-dimensional Bertotti-Robinson solution 
($AdS_2\times S^3$):
\begin{equation}\label{bert5}
ds_B^2 = \left(1-\frac{x^2}{m^2}\right)\,dt^2 +
\left(\frac{x^2}{m^2}-1\right)^{-1}dx^2 + m^2
\bigg[(d\eta-\cos\theta d\varphi)^2 + d\theta^2 + \sin^2\theta d\varphi^2
\bigg]\,.
\end{equation}
This has the same reduced three-dimensional metric as the Tangherlini solution
(\ref{tan2}) (with $r=x+m$), so there is a $G_2$ transformation $P_{TB}$ 
relating the two, which can be obtained as the product transformation 
Tangherlini $\to$ Schwarzschild black string $\to$ Bertotti-Robinson:
\begin{equation}
P_{TB} = P_{TS}P_{SB}\,.
\end{equation}
The $7\times7$ matrix $P_{TS}$ is
\begin{equation}
P_{TS} = \left(\begin{array}{ccc}
P_{(3)TS} & 0 & 0 \\
0 & P_{(3)ST} & 0 \\
0 & 0 & 1 
\end{array}\right)\,,
\end{equation} 
where $P_{(3)ST}$ is the matrix (\ref{P3ST}), and $P_{(3)TS} = 
P_{(3)ST}^{-1}$. To determine the matrix $P_{SB}$, we use the fact that the 
spherically symmetric Schwarzschild and Bertotti-Robinson solutions depend
on the same real potential $\sigma$, and so, from subsection \ref{sub33}, can 
be written as 
\begin{equation}
M_S = \eta_S{\rm e}^{A_S\sigma}\,, \qquad M_B = \eta_B{\rm e}^{A_B\sigma}\,,
\end{equation}
with
\begin{equation}\label{etaASB}
\eta_B = P_{SB}^T\eta_SP_{SB}\,, \qquad A_B = P_{BS}A_SP_{SB}\,.
\end{equation}
$A_S$ is diagonal, so $P_{SB}$ is the matrix which diagonalizes $A_B$ and 
satisfies the first equation (\ref{etaASB}). The resulting matrix $P_{TB}$ is
\cite{spinem5}
\begin{equation}
P_{TB} = \frac12\left(\begin{array}{ccccccc}
1 & 0 & 0 & 1 & 0 & 0 & \sqrt2 \\ 
0 & 0 & \sqrt2 & 0 & \sqrt2 & 0 & 0 \\ 
0 & \sqrt2 & 0 & 0 & 0 & -\sqrt2 & 0 \\ 
1 & 0 & 0 & 1 & 0 & 0 & -\sqrt2 \\ 
0 & \sqrt2 & 0 & 0 & 0 & \sqrt2 & 0 \\ 
0 & 0 & -\sqrt2 & 0 & \sqrt2 & 0 & 0 \\ 
\sqrt2 & 0 & 0 & -\sqrt2 & 0 & 0 & 0 \\ 
\end{array}\right)\,.
\end{equation}

As in the case of EM4, the rotation-generating transformation is the product
\begin{equation}
\Sigma(\Omega,\gamma) = P_{BT}
{\cal R}(\Omega,\gamma)P_{TB}\,,
\end{equation}
where ${\cal R}(\Omega,\gamma)$ is the coordinate 
transformation (\ref{R}). By construction, the transformation 
$\Sigma(\Omega,\gamma)$ acting on an asymptotically
Tangherlini solution generates an asymptotically Myers-Perry solution.
We have applied this transformation to the (singular)
static black ring of \cite{ER02a}. This complex procedure, involving repeated 
reductions (dualizations) and oxidations (inverse dualizations) according to 
the diagram
$$
\begin{array}{ccccccc}
& & & {\cal R}(\Omega, \gamma) & & & \\
ds^2 & & d\hat{s}^2,\,\hat{A} &  \longrightarrow & 
d\hat{s}^{\prime2},\,\hat{A}' &  & ds^{\prime2},\,A'\\
\downarrow & & \uparrow & & \downarrow & & \uparrow \\
{\cal M} & \rightarrow & \hat{\cal M} & & \hat{\cal M}' & \rightarrow &
{\cal M}' \\ 
& P_{TB} & & & & P_{BT} &   
\end{array}
$$
leads \cite{spinem5} to a complicated solution describing a rotating black 
ring with multipole electric and magnetic moments. A drawback of this 
procedure is that, as in the case of EM4, it transforms horizons into 
horizons, and singularities into singularities, so that our rotating black
rings are singular. 

\subsection{Other recent developments}
In \cite{GLP}, Gaiotto, Li and Padi constructed a class of multicenter 
solutions of EM5 as geodesics
\begin{equation}
M = \eta{\rm e}^{A\sigma}\,,
\end{equation}
with $\sigma$ a harmonic function and 
\begin{equation}\label{A30}
A^3 = 0\,,
\end{equation}
implying, for $A \in g_2$, Tr$(A^2) = 0$. This certainly does not
exhaust the subject, as EM5 can be consistently truncated to E5 which, as we 
have seen in subsection \ref{sub33}, admits three classes of multicenter 
solutions. The general analysis remains to be done.

In \cite{GS08}, Gal'tsov and Scherbluk discussed the hidden symmetries of 
five-dimensional supergravity with three Abelian gauge fields, a truncated 
toroidal compactification of $D = 11$ supergravity, and showed that reduction
of this theory to three dimensions leads to the $SO(4,4)/SO(2,2)\times 
SO(2,2)$ gravitating sigma model. They constructed coset representatives as 
real $8\times8$ matrices and, as an application, derived the doubly rotating 
black hole solution with three independent charges.    

By identification of the three vector fields, this $U(1)^3$ five-dimensional 
supergravity can be contracted to minimal five-dimensional supergravity which,
as we have seen, admits upon reduction to three dimensions the invariance 
group $G_{2,(2)} \subset SO(4,4)$. Contraction of the matrix representatives of
the $U(1)^3$ theory thus leads to an $8\times8$ matrix representation 
\cite{GS08} of the $G_{2(2)}/SL(2,R) \times SL(2,R)$ coset, alternate to the
$7\times7$ representation given above. It is not clear which is more simple to
use for solution generation.

\subsection{Summary}
We have seen that EM5 reduced to three dimensions leads to the 
$G_{2(2)}/SL(3,R)\times SL(2,R)$ gravitating sigma model. The first 
applications to the generation of charged solutions from neutral solutions 
could be generalized by extending the Giusto-Saxena approach. We have also 
discussed the generation of rotating solutions from static solutions via the 
five-dimensional Bertotti-Robinson solution.

\vskip1cm
\subsection*{Acknowledgments}
I am grateful to the organisers of the Bremen meeting for the opportunity
to present this review. I also wish to thank my collaborators, most
specially Dmitry Gal'tsov, Cedric Leygnac, and Adel Bouchareb.

\end{document}